\documentclass[12pt,preprint]{aastex}    

\def\arcsec{$^{\prime\prime}$}
\def\ergssec   {~ergs~s$^{-1}$}
\def\kms   {~km~s$^{-1}$}

\input{psfig.sty}

\shorttitle{NGC 4472 Hot Gas}
\shortauthors{Biller et al.}

\begin{document}           

\title{Hot Gas Structure in the Elliptical Galaxy NGC 4472}

\author{B.A. Biller\altaffilmark{1}}
\affil{Steward Observatory, 933 N. Cherry Ave., Tucson, AZ 85721}

\author{C. Jones, W. R. Forman, and R. Kraft}
\affil{Harvard-Smithsonian Center for
Astrophysics, 60 Garden St., MS-21, Cambridge, MA 02138}  


\author{T. Ensslin}
\affil{Max-Planck-Institut f\"ur Astrophysik}
\altaffiltext{1}{previously at Harvard-Smithsonian Center for Astrophysics,
email: bbiller@as.arizona.edu}


\begin{abstract} 

We present X-ray spectroscopic and morphological analyses using
Chandra ACIS and ROSAT observations of the giant elliptical galaxy NGC
4472 in the Virgo cluster.  We discuss previously unobserved X-ray
structures within the extended galactic corona.  In the inner $2'$ of
the galaxy, we find X-ray holes or cavities with radii of $\sim$2 kpc,
corresponding to the position of radio lobes.  These holes were
produced during a period of nuclear activity that began $1.2 \times
10^7$ years ago and may be ongoing.  We also find an asymmetrical edge
in the galaxy X-ray emission $3'$ (14 kpc) northeast of the core and
an $\sim8'$ tail (36 kpc) extending southwest of the galaxy.  These
two features probably result from the interaction of NGC 4472 gas with
the Virgo gas, which produces compression in the direction of NGC
4472's infall and an extended tail from ram pressure stripping.
Assuming the tail is in pressure equilibrium with the surrounding gas,
we compute its angle to our line of sight and estimate that its true
extent exceeds 100 kpc.  Finally, in addition to emission from the
nucleus (first detected by Soldatenkov, Vikhlinin \& Pavlinsky), we
detect two small extended sources within $10''$ of the nucleus of the
galaxy, both of which have luminosities of $\sim7\times10^{38}$ ergs
s$^{-1}$.

\end{abstract}

\section{Introduction \label{sec:intro}}

Significant emission from $\sim$1 keV thermal plasma coronae is common
in luminous elliptical galaxies.  This coronal gas was first detected
with the Einstein observatory \citep[]{einstein}.  Later, BBXRT,
ROSAT, and ASCA (eg. \citet[]{cor1, cor2, ROSAT, cor4, asca}) provided
improved measurements of the temperature structure and the first
information on heavy element abundances in the gas.

The morphologies of these hot coronae in bright ellipticals differ
greatly from the stellar morphologies.  While the stellar luminosity
in bright ellipticals generally follows a symmetric de Vaucouleurs
profile, the X-ray emission is often quite asymmetrical and varies in
structure from galaxy to galaxy.  Typical structures include X-ray
cavities generally corresponding to radio lobes e.g. M84 
\citep[]{M84}), sharp edges in brightness e.g. N507  \citep[]{N507}
and NGC 1404 \citep[]{N1404}) , a disk in NGC 1700
\citep[]{NGC1700}, extended bright ``arms'' (as in NGC 4636
\citet[]{N4636}), and the very extended tail seen in M86
\citep[]{M86}.  Since these structures owe their origins primarily to
either nuclear outbursts or interactions with other galaxies or the
surrounding cluster gas, X-ray observations provide a unique
opportunity to study both the histories of bright cluster elliptical
galaxies and their interactions with their environment.

Coronal gas structures can be divided into two primary categories: 1)
those produced by internal processes and 2) those produced by
galaxy-galaxy and galaxy-cluster interactions.  Nuclear outbursts are
the primary internal stimuli that shape the X-ray coronae of many
elliptical galaxies.  During periods of nuclear activity, relativistic
jets can inject $10^{52-58}$ ergs into the hot ISM, producing buoyant
radio-emitting lobes.  As they form and rise, these lobes expand, thus
both evacuating and compressing the surrounding X-ray gas, resulting in
bright-rimmed X-ray cavities coincident with the radio lobes.  Such
cavities associated with radio sources have been observed in M84
\citep[]{M84}, in NGC 1275 \citep[]{N1275b, N1275a}, and in the central
cD galaxy of the clusters Hydra A \citep[]{Hydra}, Abell 2052
\citep[]{A2052}, and Abell 478 \citep[]{A478}.  Similar X-ray cavities
without corresponding radio sources have also been observed in NGC
4636 \citep[]{N4636} and in the central galaxies of Abell 2597
\citep[]{A2597} and Abell 4059 \citep[]{A4059}.  These ``ghost''
cavities suggest that X-ray cavities persist even after the
corresponding radio lobes are too faint to be detected.  Calculating
ages for X-ray cavities/radio lobes from the buoyant rise time provides
valuable information about the nuclear outburst histories of bright
elliptical galaxies \citep[]{bubblemod}.

The interaction of bright elliptical galaxies with other galaxies and
the cluster medium shapes other structures within the coronal gas.
Interactions mold galactic gas in two ways -- 1) through ram pressure
as the galaxy passes through a gas-rich medium or 2) through
gravitational tidal forces.  Ram pressure stripping can occur both as
a galaxy passes through cluster gas and during galaxy-galaxy
interactions (e.g. Nulsen 1982 \citep[]{M86_dust, M86}.  Typically,
ram pressure  produces compression of the gas (and
corresponding steep changes in X-ray brightness by a factor of 2-10 --
see \citet[]{M84} and \citet[]{N1404b}) in the direction of motion and
elongation of the corona in the direction opposite the motion.  This
creates an overall ``cometary'' profile in the galaxy gas.  Tidal
forces due to gravity also can produce elongations and sharp edges
\citep[]{grav_model}.  In the case of galaxy-galaxy interactions,
these structures trace the merger history of the galaxies.

Coronal gas was first observed in the giant elliptical NGC 4472 with
Einstein \citep[]{einstein, einstein2}.  With ROSAT, \citet[]{cor2} 
measured the radial profile of the gas temperature and iron
abundance, while \citet[]{ROSAT} observed an elongation towards the SW
and a compression to the NE.
\citet[]{ROSAT} attributed these structures in the gas to ram pressure from
NGC~4472's passage through the Virgo ICM.  Recently, Soldatenkov,
Vikhlinin \& Pavlinsky (2003) reported detection of the nucleus of
NGC4472 in X-rays and showed that this emission is soft.  In this
paper, we report on previously unresolved structure within the
galaxy's coronal gas.  In addition to features related to those
reported by \citet[]{ROSAT}, we find X-ray cavities associated
with radio lobes within the inner $2'$ of the galaxy and two luminous
($\sim7\times10^{38}$ \ergssec) extended sources near the nucleus of
the galaxy.

This article explores the roles of nuclear outbursts, tidal
deformation, and
ram-pressure in forming structures in the X-ray coronal gas
of NGC 4472.  In \S\ref{sec:analysis}, the X-ray analysis is described.
The overall morphology of NGC 4472 is explored in \S\ref{sec:morph}.  We
discuss structures produced by nuclear outbursts in \S\ref{sec:lobes}.
Structures resulting from ram pressure and/or tidal
deformation are 
discussed in \S\ref{sec:largescale}.  The two small extended 
sources found within $10''$ of the nucleus and the nuclear emission itself
are discussed in \S\ref{sec:inner}.  Summary and conclusions are 
provided in \S\ref{sec:conc}.

\section{Analysis of X-Ray Observations \label{sec:analysis}}
NGC 4472 was observed for 41.6 ksec on 2000 June 12 with the 
ACIS S-array in very faint (vf) 
timed exposure mode (accepting ASCA grades 0,2,3,4, and 6).  The S1-S4 
and I2-I3 CCDs were operated at a focal plane temperature of -120 C
and with a frame time of 3.214 s.  The Chandra pointing placed the center of
the galaxy on the S3 CCD.  We extracted useful NGC4472 data from the 
S2 and S3 CCDs.  We used data from the S1 and I2 CCDs for background 
determination.  At this time in the Chandra mission, the contamination on the 
ACIS filter (resulting in decreased low energy efficiency) was fairly modest.

We time filtered the S2 and S3 CCDs independently for background
flares following the procedures from \citet[]{cleaning}.  Lightcurves
were created with a binsize of 259.28 s (this corresponds to 80 3.214
s frames) using the lc\_clean software.  We found a mean count rate of
0.175 cts/s in the energy band from 0.3-10 keV and excluded 1987 s for
S2 (when cts/s were $\geq 3\sigma$ above the mean count rate).  A
total clean exposure time on S2 of 40.2 ksec was retained.  Since the
large angular extent of the galaxy made background light curve
determination difficult for the back-illuminated S3 CCD, we derived
background light curves from the back-illuminated S1 CCD.  The S1 CCD
possesses a response similar to that of the S3 CCD.  For S1, we found
a mean count rate of 1.748 cts/s in the energy band from 0.3-10 keV
and excluded 7428 s due to high background intervals, leaving a total
clean exposure time of 34.7 ksec on S3.  Bright columns and pixels due
to instrumental effects or cosmic ray afterglows also were removed.

Since we are interested in the diffuse gas emission from NGC 4472, we 
identified and excluded regions around 214 point source candidates
with 3 or more counts in the 0.5-2 keV energy band.  
Point source candidates were identified by
performing a wavelet deconvolution on the data.  
We also created and examined a hard image
in the 2-5 keV band to detect any highly absorbed sources.  No such
sources were found.  

\section{Overall Morphological Features \label{sec:morph}}

\begin{figure}
\plotone{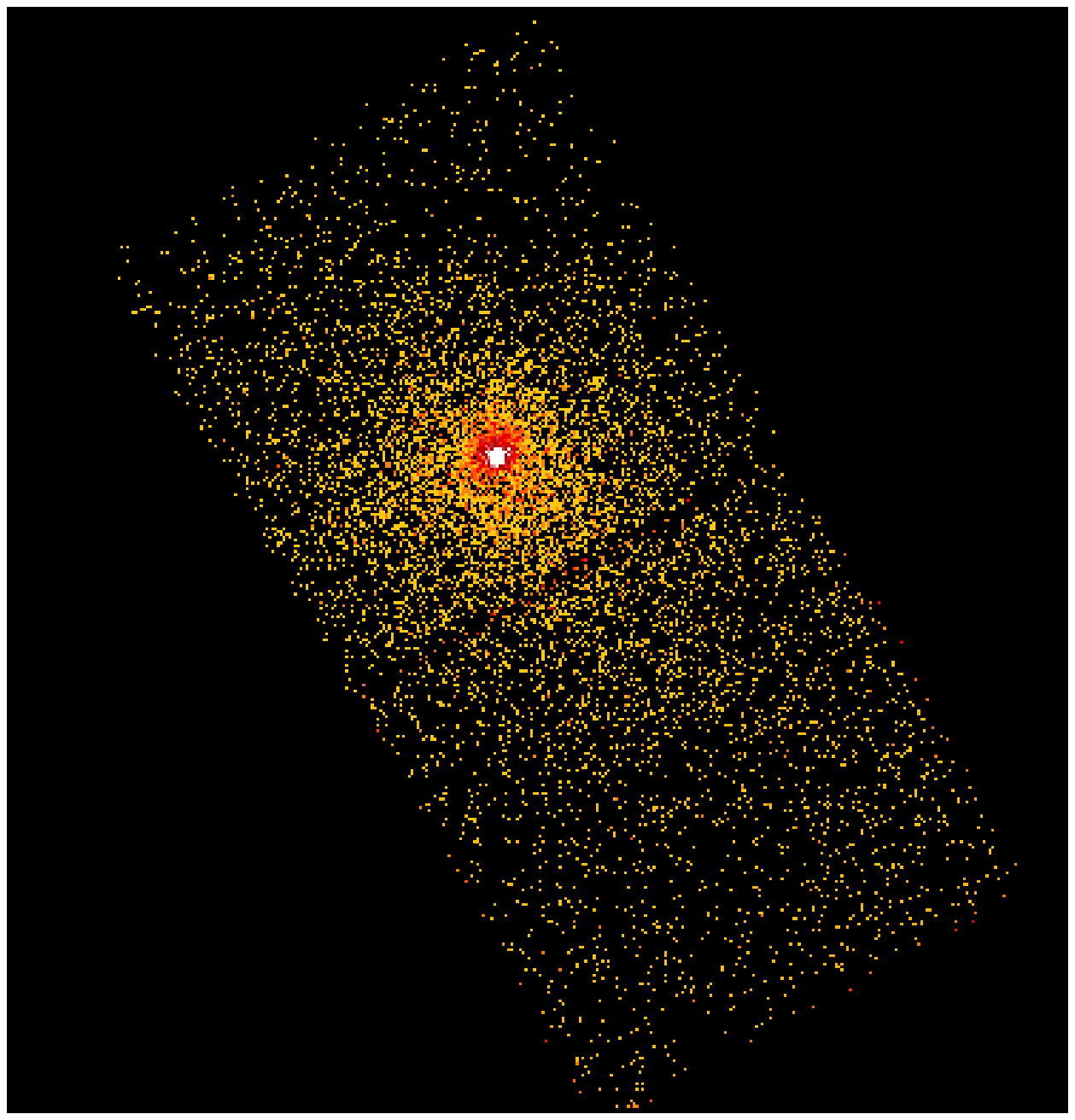}
\caption{Chandra NGC 4472 bin 2 image in the 0.7-2.0 keV band.
The side of each chip spans 8.3 arcmin.  At the distance of
Virgo, 1 arcmin corresponds to a linear size of 4.65 kpc.  North is
to the top of the image.}
\label{fig:images}
\end{figure}

\begin{figure}
\plotone{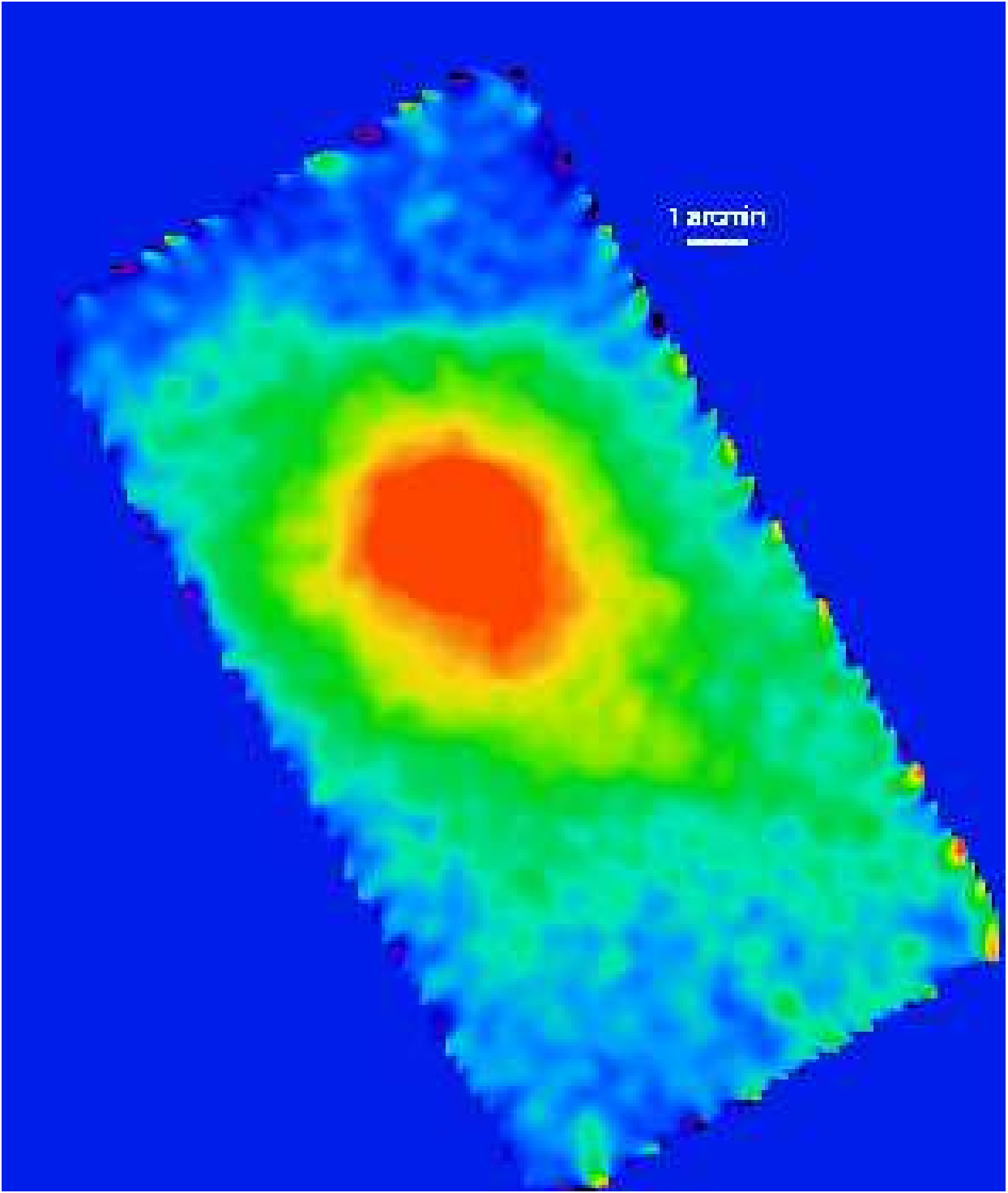}
\caption{Smoothed exposure map corrected Chandra image of NGC 4472 in
the 0.7-2 keV band.  At the distance of
Virgo, 1 arcmin corresponds to a linear size of 4.65 kpc.  North is
to the top of the image.}
\label{fig:expimg}
\end{figure}

Fig.~\ref{fig:images} shows NGC 4472 on the S2 and S3 CCDs in the
energy range 0.7-2 keV.  This image was produced from the ``cleaned''
events, with a binning factor of 2 (1 binned pixel is 2$\times$0.492
arcsec).  A background subtracted, exposure map corrected image is
presented in Fig.~\ref{fig:expimg}.  This image has been smoothed on a
length scale of 5 pixels.  At the distance of NGC 4472 (16 Mpc --
\citet[]{distance}), each arcmin corresponds to a linear size of 4.65
kpc.

We characterize the gas morphology into four primary features.  
A bright, symmetric 
central region dominates the emission.  Within the inner $2'$ (9.3 kpc)
of this region are radio lobes corresponding to X-ray cavities
(see \S\ref{sec:lobes}).  On larger scales, a drop in surface brightness 
is apparent towards the northeast, three
arcminutes (14 kpc) north of the nucleus.
Finally, southwest of the nucleus, a tail of emission stretches 
nearly $8'$ (37 kpc) from
the nucleus and becomes as narrow as 10$''$ (775 pc).  (See 
\S\ref{sec:largescale})
The hot coronal gas in NGC 4472 is
actively being sculpted by both internal (nuclear outbursts) and
external (interaction with Virgo cluster gas) processes.

\section{X-Ray Structure Related to Radio Lobes \label{sec:lobes}}

\begin{figure}
\epsscale{0.7}
\plotone{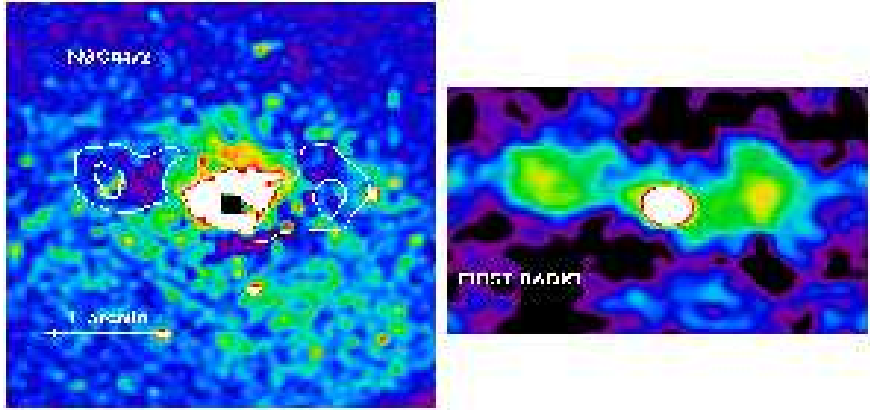}
\caption{Radio emission (right plot) from the FIRST survey and
X-ray emission in the 0.5-2 keV band 
with a smoothed radial profile subtracted out (left plot).  The radio contours 
have been overlaid on the X-ray emission.}  
\label{fig:lobes}
\end{figure}

NGC 4472 possesses weak radio lobes from an episode of nuclear
activity that is possibly still ongoing.
A FIRST radio image \citep[]{FIRST} of the inner two arcmin is shown in
Fig.~\ref{fig:lobes}.  The corresponding Chandra X-ray image (with 
a smoothed radial profile of the inner region subtracted to enhance
the asymmetric emission) is also presented
in Fig.~\ref{fig:lobes} and shows depletion or ``holes'' 
in the X-ray emission at the positions 
of the radio lobes.  Emission in the X-ray holes is depleted by $\sim$95$\%$ 
relative to X-ray emission at a similar radius from the nucleus.

We performed spectral fits in XSPEC for the lobe regions shown in
Fig.~\ref{fig:lobes} and also for an elliptical annulus $5''$
($\sim$400 pc) in width framing each lobe region.  Since the holes are
substantially depleted of X-ray gas, spectra of the lobe regions
measure primarily the overlying gas.  ACIS sky backgrounds
\citep[]{cleaning} were subtracted from the spectra.  A mekal model
with a constant neutral hydrogen column density of
1.7$\times$10$^{20}$ cm$^{-2}$ \citep{nH} was fit in the 0.5-7 keV
band.  We found a temperature of $0.82\pm0.02$ keV for lobe 1 (east
and to the left in Fig.~\ref{fig:lobes}) and and a similar temperature
of $0.76\pm0.03$ keV for lobe 2 (west and to the right in
Fig.~\ref{fig:lobes}).  (Errors correspond to 90$\%$ uncertainties.)
The gas temperature dropped slightly to $0.77\pm0.02$ keV in a $5''$
elliptical annulus outside the eastern lobe and rose slightly in a
similar annulus outside the western lobe to 0.86$^{+0.02}_{-0.01}$
keV.  Hence, we see no evidence for significant shock heating along
the edges of the lobes.
  
Assuming the lobes lie in the plane of the sky,
we estimate a minimum energy and age for the two X-ray cavities/radio lobes.  
The lower bound on energy is simply the
mechanical energy necessary to evacuate the X-ray gas from the cavity
and is estimated from the pressure of the surrounding gas and the
volume of each cavity.  Parameters for each lobe are given in
Table~\ref{tab:lobes}.  The initial electron density (n$_e$) was taken from our
$\beta$ model fit to the surface brightness at the projected radius of
the lobe (see \S\ref{sec:specmorph}), kT from our temperature fits for
each lobe, and the volume (V) from the lobe size.  Lobes were
considered ellipsoids -- the radius along the line of sight was
modeled as the apparent semi-minor axis of the elliptical spectral
extraction region on the sky.  Estimating a distance from the nucleus of
3.6 kpc and a sound speed of 300 km s$^{-1}$ (appropriate for a 0.8
keV gas), and assuming a buoyant lobe moves nearly at the sound
speed, the lobes were produced by a nuclear outburst that began
at least $\sim$1.2$\times$10$^7$ years ago.  We also utilized code from
\citet[]{bubblemod} to model the rise velocity as the lobe buoyantly
rises through the halo gas.  We find a lobe velocity from these
models of $\sim$320 \kms, approximately the sound speed.

With a mechanical evacuation energy of $\sim$10$^{54}$ erg ($E$ from
Table~\ref{tab:lobes}), since the lobes likely contain relativistic
plasma,  we can estimate the enthalpy of the lobes as
$4\times E$, or about $4 \times 10^{54}$ erg \citep[]{syscav}.  Thus,
the energy in the X-ray cavities in NGC 4472 is small compared to those in
other galaxies, which possess energy outputs of up to 10$^{61}$ ergs
\citep[]{A2597}.  While the energy in the lobes is
sufficient to evacuate a cavity in the X-ray gas, it does not produce
bright rims, as seen in other systems.

\begin{table}
\caption{Parameters for Radio lobes \label{tab:lobes}}
\begin{tabular}{ccccc} \hline \hline
Lobe & n$_e$ (cm$^{-3}$)$^*$ & kT (keV)$^*$ & V (kpc$^3$) & E
(10$^{53}$ergs)\\ \hline
East & 0.059 & 0.82 & 23 & 5.2 \\ 
West & 0.080 & 0.76 & 27 & 7.9 \\ \hline
\end{tabular}

$^*$ ambient gas density and pressure
\end{table}

\section{Large Scale Structures in the X-ray Coronal Gas \label{sec:largescale}}

\subsection{Analysis of Temperature Structure and Morphology \label{sec:specmorph}}

\begin{figure}
\plotone{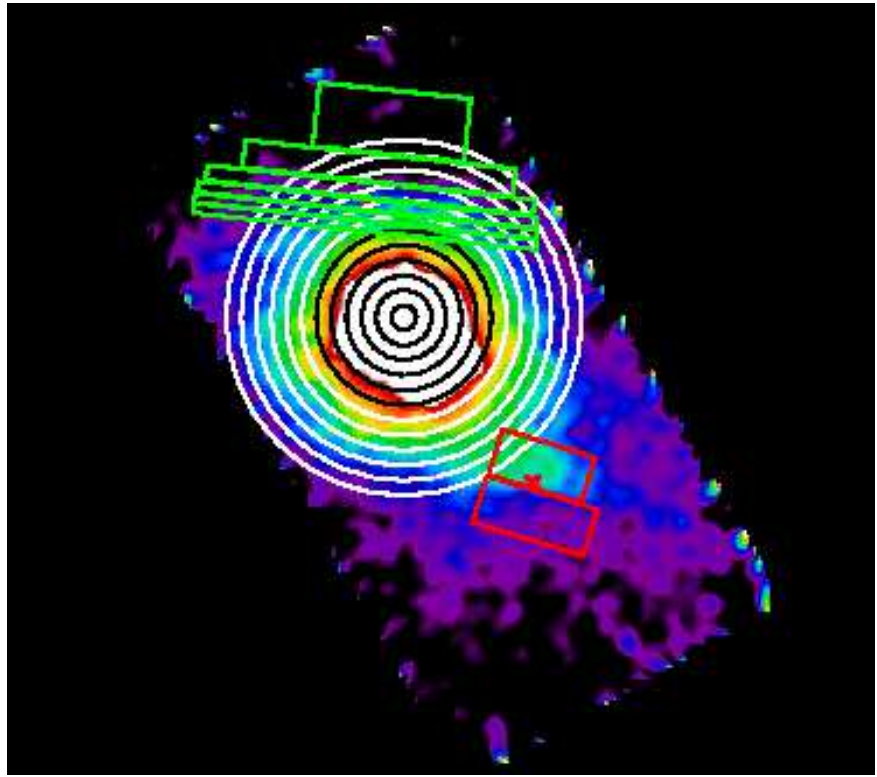}
\caption{Regions used for spectral extraction overlaid on the smoothed
exposure map corrected image.  Red crosses mark the position of
arbitrary points used for projected distances in the NE edge and SW
tail.  North is towards the top of the image.}
\label{fig:regs}
\end{figure}

\begin{figure}
\epsscale{0.7}
\plotone{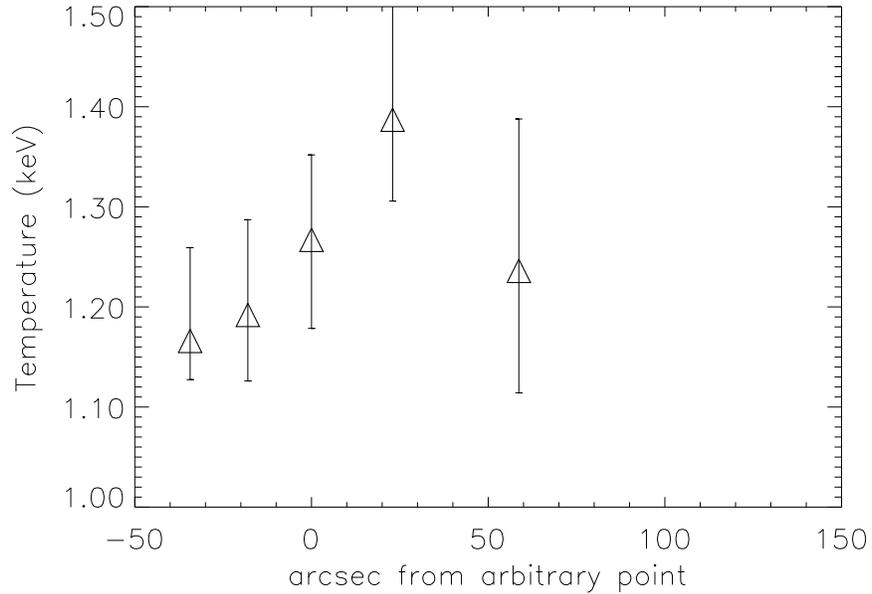}
\caption{Projected X-ray temperature profile for the NE edge.
Distances are projected from an arbitrary point shown in
Fig.~\ref{fig:regs} and perpendicular from the long edge of the
spectral extraction regions.}
\label{fig:north}
\end{figure}

We investigated the temperature structure in the central region,
across the northern edge, and in the tail.  In the central region,
spectra were extracted for ten annular regions to a $4'$ (19 kpc)
radius from the nucleus.  An image with these spectral extraction
regions overlaid is shown in Fig.~\ref{fig:regs}.  We subtracted ACIS
sky backgrounds \citep[]{cleaning} from our spectra.  Spectra were fit
in the 0.7-2 keV band using a mekal model in XSPEC.  Spectra were
extracted on the S3 CCD only -- sections of each annulus outside S3 were
excluded.  The neutral hydrogen column density was fixed at
1.7$\times$10$^{20}$ cm$^{-2}$ \citep{nH}.  We find that the
temperature increases monotonically from 0.8$\pm$0.05 keV to
1.1$\pm$0.05 keV radially outward to 10 kpc, then rises by 20$\%$ to
1.2$\pm$0.1 keV from 10-12 kpc, and returns to 1.1$\pm$0.05 keV at
radii greater than 12 kpc.  The 10 kpc temperature jump coincides with the
bright NE edge and is also apparent in spectra extracted along this
edge (Fig.~\ref{fig:north} -- see below for details.)

\begin{figure}
\plotone{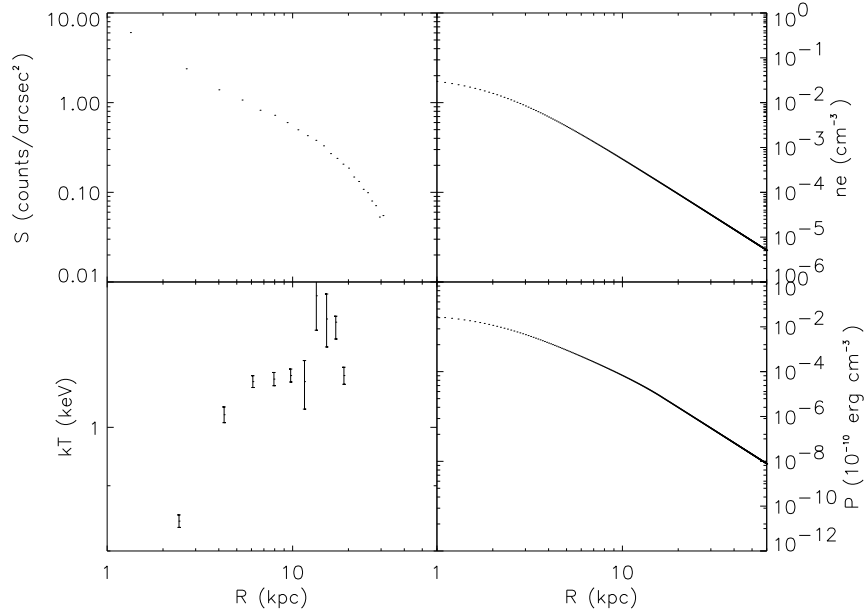}
\caption{Radial surface brightness (upper left), X-ray temperature
(lower left), electron density (upper right), and pressure (lower
right) profiles for the inner
$\sim$4 arcmin of the galaxy.   Pressure and density profiles are
derived from our $\beta$ model fit to the surface brightness profile.}
\label{fig:center}
\end{figure}

After applying an exposure map correction, subtracting background
counts, and excluding point sources (as well as the extended central
sources, see \S\ref{sec:inner}) we also measured the surface
brightness distribution in 22 annuli covering the same $4'$ (19 kpc)
radius region.  The surface brightness distribution is strongly peaked
around the nucleus of the galaxy and decreases with radius.  An
increase in surface brightness as observed here, along with a decrease
in temperature, is consistent with a cooling flow.  We fit a $\beta$
model to our surface brightness profile and found $\beta$ =
0.440$\pm$0.001 and a core radius a = 0.280$\pm$0.001 kpc.  Density
and pressure were derived from this model and also are plotted as a
function of radius in Fig.~\ref{fig:center}.  Adopting a central
temperature of 0.8 keV and a central density of 0.03 cm$^{-3}$ from
our $\beta$ model fit, we calculated a cooling time at the galactic
center of $\sim$0.9 Gyr, consistent with a cooling flow.

Although the northeast edge shows a drop in surface brightness by an
order of magnitude, this is not apparent in an azimuthally averaged
radial surface brightness profile.  To directly investigate this
surface brightness drop, we considered both temperature and surface
brightness locally along the NE edge.  Surface brightnesses were
calculated in a 50$^{\circ}$ wide wedge centered on the northern edge
after an exposure map correction had been applied.  This surface
brightness profile is presented in Fig.~\ref{fig:N4472prof}.  The
surface brightness decreases by an order of magnitude over a length of
$50''$ (3.9 kpc).  Beyond $100''$ (7.8 kpc) of the edge, hot cluster
gas dominates over the galactic gas.

In order to examine the temperature structure of the NE edge, we
extracted spectra for a series of rectangular slices with long sides
aligned parallel to the edge.  Slices were chosen to contain at
least 1000 source counts.  An image with these spectral extraction
regions overlaid for the NE edge is shown in Fig.~\ref{fig:regs}.
Along the NE edge, emission from Virgo cluster gas comprises a
significant portion (up to 20$\%$) of the total emission.  We fit a
mekal model to the cluster gas far from the galaxy (using the I2 CCD)
and found a temperature of 2.2$_{-0.5}^{+1.6}$ keV.  We used this
cluster model as a second mekal model in fitting our NE edge spectra.
Spectral fits were performed in XSPEC in the 0.7-2 keV band.  A
temperature profile for the NE edge is plotted in
Fig.~\ref{fig:north}.  For the NE edge, the temperature increases from
1.1$\pm$0.1 to 1.4$\pm$0.1 keV towards the north.

\begin{figure}
\plotone{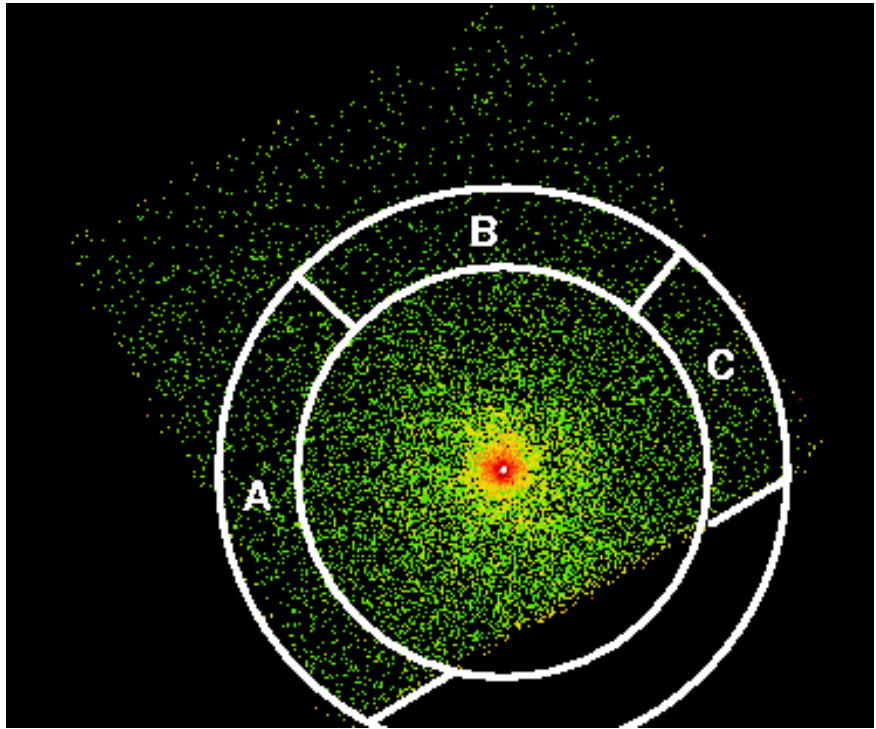}
\caption{Spectral extraction regions for annular wedges along the NE
edge.}
\label{fig:wedges}
\end{figure}

To determine the extent of the hotter emission along the northern edge
of the galaxy, we fit three wedges displayed in Fig.~\ref{fig:wedges}
(labeled A, B, and C).  Each wedge had an inner radius of $2.8'$ (13
kpc) and an outer radius of $4.1'$ (19 kpc) arcmin from the nucleus.
We find temperatures of $1.11\pm0.01$ keV for wedge A, $1.40\pm0.04$
for wedge B, and $1.37\pm0.04$ for wedge C.  Thus, the hotter region
is north and east of the galaxy center.  This temperature change is
likely produced by the compression of the corona as the galaxy moves
through the Virgo ICM.

\begin{figure}
\plotone{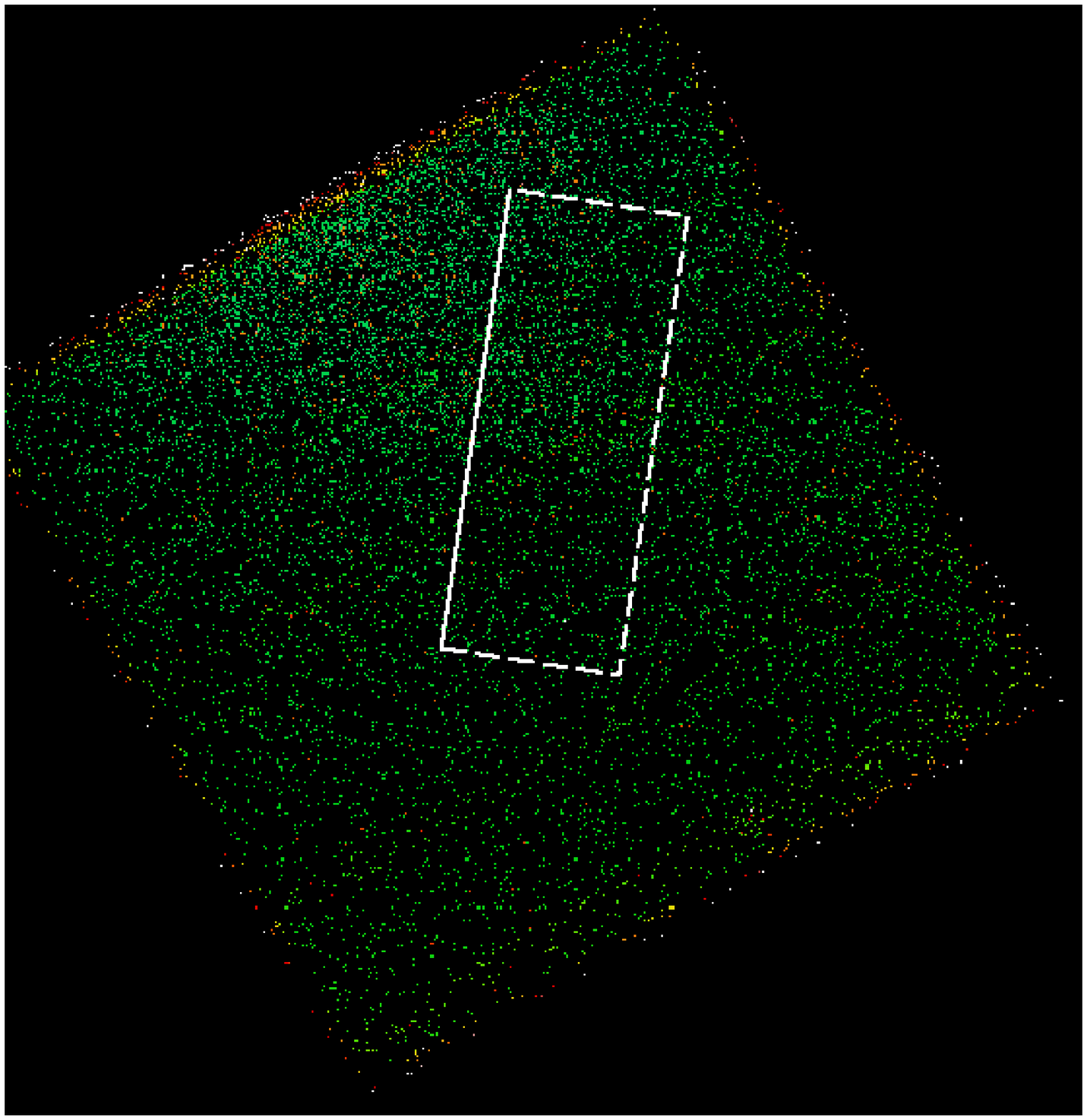}
\plotone{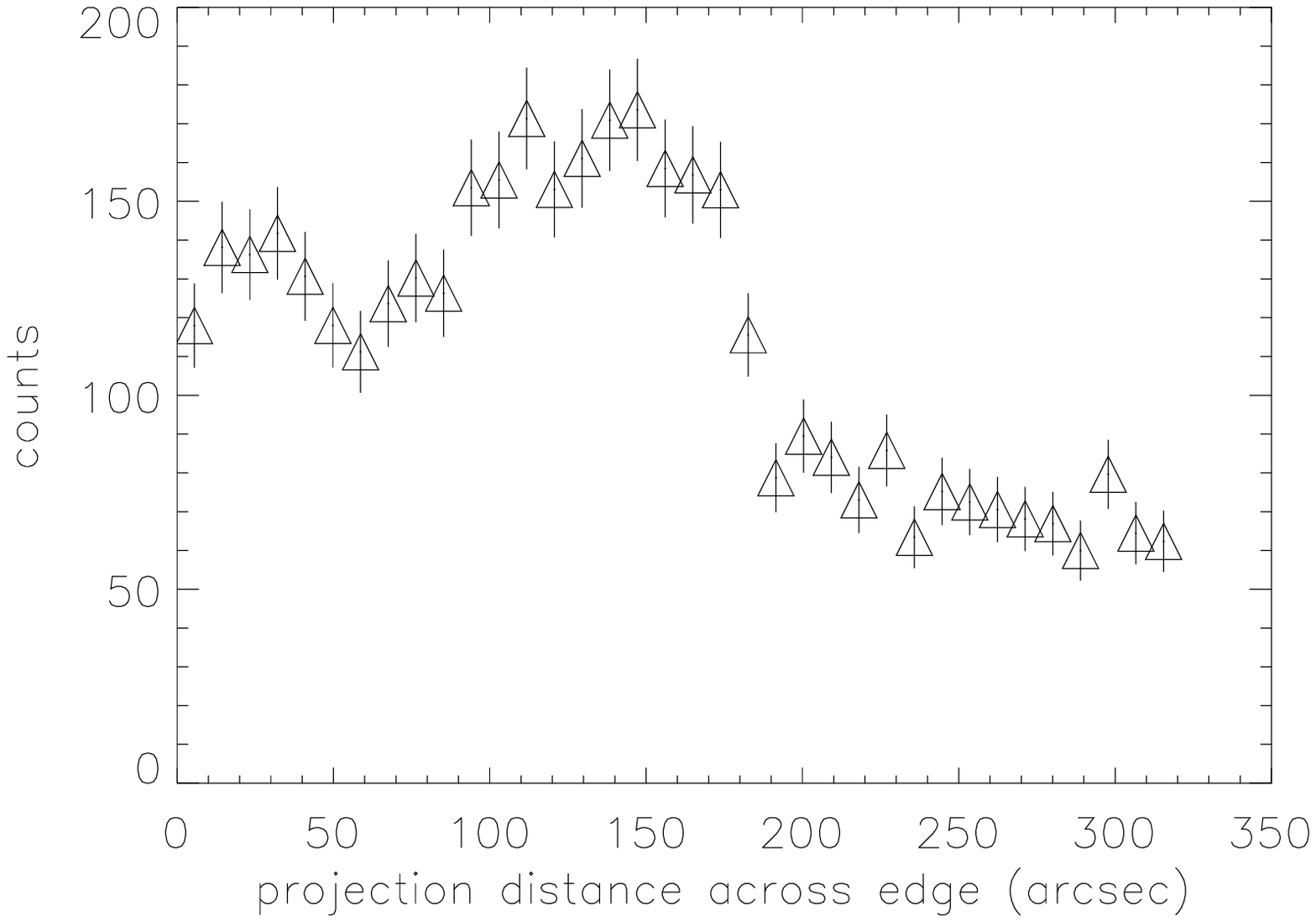}
\caption{Surface brightness extraction region (top) and surface
brightness profile (bottom) for the SW tail of the galaxy.
The positive direction is SE, away from the galaxy core. }
\label{fig:south}
\end{figure}

We extracted and fit temperatures on and off the SW tail.
Spectral extraction regions along the SW tail are shown in
Fig.~\ref{fig:regs} and were chosen to contain at
least 1000 source counts.  To include the contribution of
cluster gas in our spectra, backgrounds derived far 
from the galaxy (using the I2 CCD) were
subtracted from these spectra.  A mekal model was fit to each
spectra.  We find a temperature of $1.10\pm0.02$ on the
tail and a temperature of $1.35\pm0.06$ south of the
tail.  Surface brightnesses
were calculated for a series of $7.9''\times113.4''$ slices across the SW tail
after an exposure map correction had
been applied.  The surface brightness across the tail is plotted in
Fig.~\ref{fig:south}; the positive direction is to the SE, away from 
the galaxy core.  The region from which the surface brightness
profile was extracted is also plotted in Fig.~\ref{fig:south}.
The surface
brightness peaks by a factor of three on the tail, compared to emission
from the galaxy halo at the same radius.
The change in surface brightness along the southern edge of the 
filament is sharp, occurring over an angular
size of $\leq$20$''$ (155 pc).

\subsection{Extent of Large Scale Structures}

\begin{figure}
\plotone{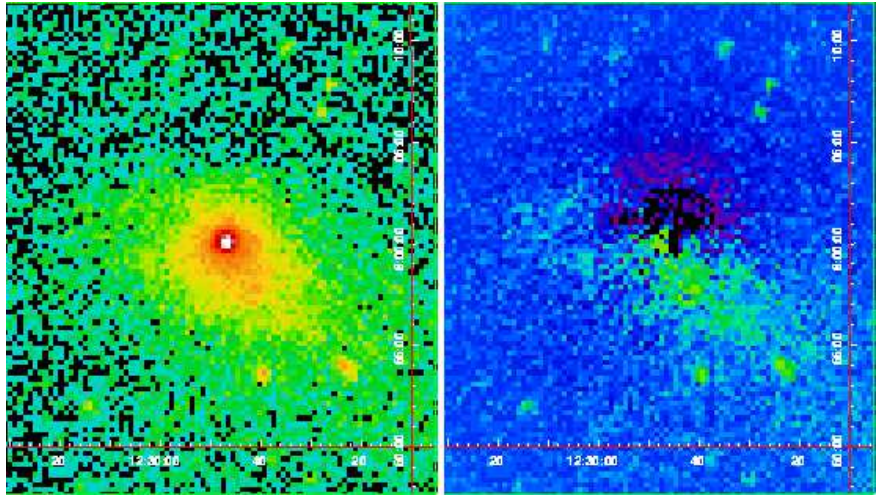}
\caption{Raw image (left) and radial profile subtracted
image (right) from ROSAT PSPC data.  Coordinates are J2000.}
\label{fig:ROSAT}
\end{figure}

To determine the extent of asymmetric features in the NGC 4472 halo
gas on large scales, we used the ROSAT PSPC pointed observation and
subtracted a symmetric radial profile from the image.  
NGC 4472 was observed by ROSAT for 25 ksec on 26 January 1992 (see
\citet[]{cor2} and \citet[]{ROSAT} for details on this observation).  
A symmetric model based on the radial surface brightness profile was
subtracted from the inner $15'$ (70 kpc) radius of the ROSAT PSPC image.  Both the 
ROSAT PSPC image and the subtracted image are presented
in Fig.~\ref{fig:ROSAT}.  A faint extension, $\sim$4$'$ (19 kpc) long, 
is seen to the ENE.  The large SW tail extends $11'$ ($\sim$51 kpc) 
from the core of the galaxy in the ROSAT image, compared to the
$8'$ extent seen in the Chandra image.  Thus the tail as
detected by ROSAT is fully contained within the Chandra S2 CCD, 
while the faint extension to the ENE is not in the Chandra field.

\subsection{The Formation of the Gas Discontinuity and the Tail by the Motion of NGC 4472 through the Virgo ICM}

\begin{figure}
\plotone{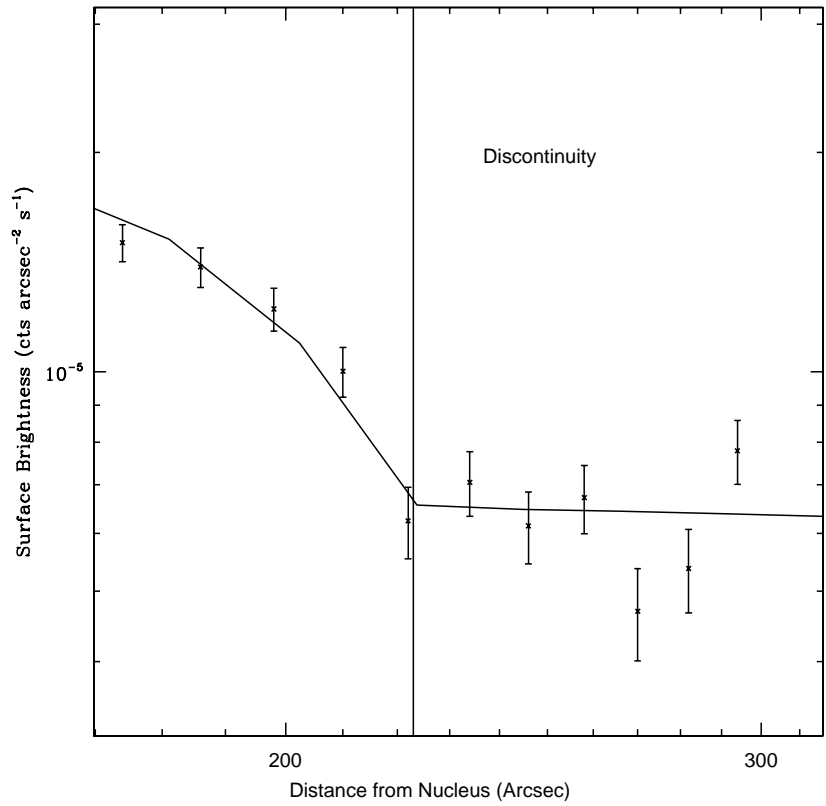}
\caption{The X-ray surface brightness measured in a wedge 50$^\circ$ wide
and centered on the northern edge is shown.  The solid curve shows
the predicted surface brightness derived from a simple model
of the density profile, as described in the text.  The vertical
line marks the approximate position of the discontinuity.}
\label{fig:N4472prof}
\end{figure}

The large-scale structure of coronal gas in NGC 4472 is formed
primarily by the interaction of the galaxy with the Virgo ICM, as
originally suggested by \citet[]{ROSAT} based on the ROSAT
observation.  The sharp surface brightness discontinuity $\sim$3.5$'$
north of the nucleus is most likely a `cold-front' which is the result
of ram-pressure as NGC 4472 falls into the Virgo cluster.  Such
structures have been well observed by Chandra in cluster mergers
(e.g. Markevitch et al. 2000, Vikhlinin et al. 2001, Markevitch et
al. 2002).  On a smaller scale, similar features have been seen in NGC
1404 in the Fornax Cluster (Dosaj et al. 2002, Machecek et al. 2004 in
preparation) and NGC 7619 \citep[]{N7619}.  Figure~\ref{fig:N4472prof}
contains a plot of the NGC~4472 surface brightness profile in a
50$^\circ$ wedge centered on the nucleus in the direction of this
discontinuity.  The appearance of the discontinuity is a result of the
temperature difference between the ISM of the galaxy
($kT\sim$1.1$\pm$0.1 keV) and that of the Virgo cluster ICM
($kT=2.2_{-0.5}^{+1.6}$ keV).  There is no evidence of shock heated
gas in the vicinity of the discontinuity implying that the infall
velocity is subsonic ($v<$ 780 km s$^{-1}$).  In order to evaluate the
`cold-front' hypothesis, we have created a simple model of the density
distribution in the region around the discontinuity.  The region just
beyond the discontinuity must be a stagnation point and the gas on
both sides of the discontinuity is in pressure equilibrium.  Thus the
change in temperature across the discontinuity is balanced by a
similar (inverse) jump in the gas density.  If we assume that the
density of the gas interior to the discontinuity is uniform, and that
the galaxy is falling into a uniform density medium, the resulting
surface brightness profile is shown in Figure~\ref{fig:N4472prof}
(continuous curve).  The rise in surface brightness just behind
(within $\sim$50$''$ of) the discontinuity can be explained by this
simple model.  Hydrodynamic simulations of the `cold-front' phenomenon
in merging clusters of galaxies suggest that such interactions can
create complex currents in the interior regions that transport the low
entropy, high abundance material from the center to just behind the
discontinuity \citep[]{cfsim}.

At first glance, the SW tail appears morphologically similar to tidal
features produced in N-Body simulations of galaxy mergers.  However,
NGC 4472 is a fairly isolated galaxy -- at least from other galaxies
massive enough to produce such a large feature.  While stellar
kinematics are slightly perturbed in the core of NGC 4472
\citep[]{star_orbit}, this perturbation is consistent with minor rather than major
merging.  NGC 4472 has most likely consumed a number of smaller
galaxies over time (e.g. UGC 7636; see
Patterson $\&$ Thuan 1992 and McNamara et al. 1994).  However, if it
had interacted with a galaxy large enough to produce the observed SW
tail, we would detect this other very massive (and bright) galaxy
somewhere in the neighborhood.

We suggest that the SW tail is a long tail of gas extending primarily
along our line of sight (similar to simulation results by
\citet[]{ram_model}).  The long pathlength of the tail produces the
apparent sharpness along the edges.  The low temperature along the
tail (1.1 keV compared to 1.4 keV in the surrounding gas) implies that
the tail was drawn out from the lower temperature galaxy gas
surrounding the core of the galaxy.  NGC 4472 has a radial velocity of
997 km s$^{-1}$, while M87 has a radial velocity of 1292 km s$^{-1}$
\citep[]{velocities}.  This radial velocity difference between NGC
4472 and M87 suggests that NGC 4472 lies behind M87 (which sits at the
center of the cluster) and is falling in towards the cluster center
and towards us partially along our line of sight.  As there is no
evidence of shock heating in the SW tail gas, we expect the tail gas
to be nearly in pressure equilibrium with the surrounding gas.

To constrain the length of the SW tail along the line of
sight, we assumed pressure equilibrium, calculated the pressure
adjacent to the SW tail, and then calculated the extension
along the line of sight that would produce a similar pressure in the SW
tail.  From our $\beta$ model fit (see
\S\ref{sec:specmorph}), we expect an electron density of 4$\times$10$^{-4}$
cm$^{-3}$ at a radius of 30 kpc just off the SW tail. 
We take a temperature of 1.4 keV for the
surrounding gas (from our off-tail spectra in \S\ref{sec:specmorph}) 
and find a pressure of 5.6$\times$10$^{-4}$ keV cm$^{-3}$.  Along the SW tail,
the gas temperature is measured as 1.1 keV as 
described in \S\ref{sec:specmorph}.  We then calculated
the pressure, volume, and density 
as a function of tail extension along the line of sight.
To achieve pressure equilibrium, the
tail must be extended $\sim$50 kpc along the line of sight at the
position of our on-tail measurement 
region.  This region lies $4.5'$ (21
kpc on the sky) from
the start of the tail near the galaxy core.  
Thus the ratio of the extension along the line of
sight to the extension in the sky is: 50 kpc/21 kpc = 2.4.  Applying this
ratio to the length of the tail on the sky, we find
a total tail length of 96
kpc for the $8'$ length observed by Chandra and 133 kpc for the $11'$
extension seen in the ROSAT PSPC image.  Modeling the tail as a cone with a 
base diameter of 5.9 kpc, we find a total tail mass of 1.05$\times$10$^7$ 
M$_{\odot}$ for the tail length observed by Chandra and a total tail mass 
of 1.45$\times$10$^7$ M$_{\odot}$ for the tail length observed by ROSAT.  For
the region of the tail on the S2 CCD, we measure a luminosity of 
2.2$\times$10$^{40}$ erg s$^{-1}$ in the 0.5 to 2 keV band.

Both the NE edge and SW tail structures are consistent with the
effects of ram pressure.  Both these structures possess sharp
edges -- interestingly, these sharp edges are produced in different manners.
The NE edge is due to the compression of the gas as NGC 4472 passes through the
Virgo ICM, while the large pathlength of the SW tail gas
produces its apparent sharpness.

\section{The Nucleus and Extended Central Sources \label{sec:inner}}

\begin{figure}
\plotone{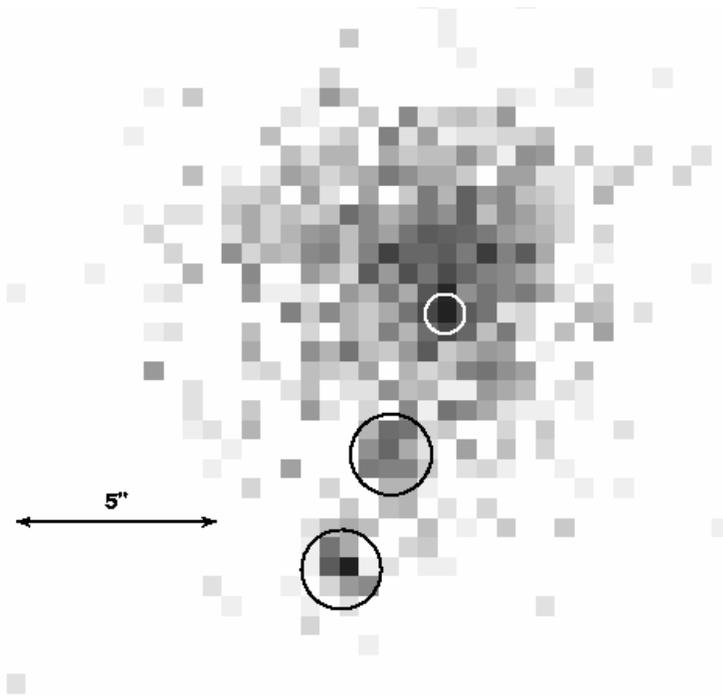}
\caption{Image of the inner $10''$ of NGC 4472.  The nucleus and the two small
extended sources are circled. The source positions and properties are
given in Table~2.
}
\label{fig:inner}
\end{figure}

Chandra's exquisite resolution reveals new features in the galaxy core -- 
emission from the nucleus itself and 
two small extended sources within $8''$ of the nucleus.
An image of the central $10''$ of NGC 4472 is presented in 
Fig.~\ref{fig:inner}.  Each of these sources contains about 200 counts 
within a $1''$ radius. 

\citet[]{4472_nuc2} and \citet[]{4472_gc} detected emission from the
nucleus at energies below 2.5 keV, although \citet[]{4472_nuc} did not
detect nuclear emission in the hard X-ray band from 2-10 keV.  Using
the accurate radio position from \citet[]{4472_nuc3}, for the nucleus,
we confirm emission in the 0.3-10 keV band.  Within $1"$ of the
nuclear position, we measure $64.0\pm19.7$ source counts in the full
ACIS-S energy band from 0.3-10 keV.  For a power-law spectrum with a
photon index of 1.7 and an n$_H$ of $1.7\times10^{20}$ cm$^{-2}$, this
corresponds to a 0.3-10 keV flux of $1.53\pm0.47\times 10^{-14}$ ergs
cm$^{-2}$ s$^{-1}$ and a luminosity of $4.74\pm1.46\times10^{38}$ erg
s$^{-1}$.  (In the 0.3-2 keV energy band, the nuclear luminosity is
$1.95\times10^{38}$ ergs s$^{-1}$.)

\citet[]{4472_gc} previously reported emission from the two sources
south of the nucleus in Fig.~\ref{fig:inner}, identified as numbers 72
and 73 in their source list. Each source has a luminosity of $\sim 7
\times 10^{38}$ \ergssec. Positions, fluxes and luminosities for these
sources are given in Table~\ref{tab:innerprop}. To determine whether
these two sources were point-like or extended, we compared the 0.5 to
2 keV surface brightness within a $1''$ circle centered on the source
to the surface brightness within an annulus from $1''$ to $2''$.
Background was computed from an annulus from $3''$
to $4''$.  We chose a small annulus for background, due to
the sharp decline in the surface brightness of NGC 4472's
corona. For comparison we also extracted the counts from five point
sources using these same annuli. The ratio of the counts from the
$1''$ circle to the counts in the $1''$ to $2''$ annulus was
$1.15\pm0.34$ for the northern source, $1.65\pm0.43$ for the southern
source and $6.6\pm1.2$ for the summed point sources. Although the bright
emission from the gas around NGC 4472 complicates the determination of
extent, these ratios demonstrate that the two inner sources are not
point-like. We note that if
the two small sources are more extended than the $3''$ to $4''$
annulus chosen for the background, the addition of source counts to
the background would have increased the ratio of source counts in the 
$1''$ circle to the counts in the $1''$ to $2''$ annulus.
 
We searched for optical and radio counterparts to the two small extended 
sources ($<$8$''$ from the nucleus) in the FIRST radio image
\citep[]{FIRST} and also in HST WFPC2 data (F814W and F555W filters).
We found no counterparts either in the optical or radio
regimes -- these sources are observed only in the X-ray. However, it is 
worthwhile to note that the large surface brightness of NGC 4472 near the 
nucleus ($\mu_B$ = 18.24 mag arcsec$^{-2}$, \citet[]{4472nuc4}) 
prevents us from detecting faint optical sources.

Because they appear to have no detectable radio or optical emission,
we first modeled the two small extended sources as X-ray emitting gas
clouds. For a simple estimate of the masses of these objects, we
modeled them as spheres with radii of $1''$ (78 pc at 16 Mpc). We
found source masses of $\sim3\times10^4$ M$_{\odot}$. To estimate
their lifetimes, we calculated cooling times for a 0.5 keV gas and a
1.0 keV gas of only $\sim$2.0$\times10^{6}$ yr or
$\sim$5.0$\times10^{6}$ yr. We also compared the expected gas pressure
within these sources ($\sim$0.5 keV cm$^{-1}$) to that of the ambient
gas (0.02 keV cm$^{-1}$).  The very short cooling time along with the
overpressured nature of these inferred gas clouds has led us to
conclude that these sources are probably not clouds of gas, but
instead may be collections of point X-ray sources.

\begin{table}
\caption{Properties of Extended Sources \label{tab:innerprop}}
\begin{tabular}{cccc}\hline \hline
Source Position  &  Net Counts$^a$ & Flux$^b$ & L$_x$ \\
 (epoch 2000)   & (erg cm$^{-2}$ s$^{-1}$) & (erg s$^{-1}$) \\ \hline
12:29:46.858 +07:59:58.26  &    $257\pm33$  &   $2.4\times10^{-14}$ &
 $7.4\times10^{38}$ \\ 
12:29:46.924 +07:59:55.28  &    $245\pm29$  &   $2.3\times10^{-14}$  &
 $7.1\times10^{38}$ \\ \hline\hline
\end{tabular}

$^a$ - net counts are in the energy band 0.3-2.0 keV
and are computed from the source counts within a $3''$ radius circle
with local background taken from a surrounding annulus.\\

$^b$ - Flux in the 0.5-2.0 keV band is computed from the net counts
assuming a 5 keV bremsstrahlung spectrum, suitable for galactic
binaries.  Note that other spectral forms, e.g., a power law spectrum
with $\Gamma=1.7$ or a thermal spectrum with $T=10^7$~K, give fluxes
that differ by less than 10\% from that for the bremsstrahlung spectrum.
\end{table}

\section{Conclusions and Summary \label{sec:conc}}

Chandra's arcsecond resolution reveals several previously
unobserved structures in the $\sim$1 keV halo gas surrounding NGC
4472.  These structures have been molded by a variety of mechanisms,
most importantly by nuclear outburst activity and ram pressure.  

On small size scales (inner $2'$ corresponding to 9.3 kpc), 
X-ray cavities corresponding to radio lobes 
are the result of an episode of nuclear activity that began 
1.2$\times$10$^7$ years
ago and injected more than $\geq10^{54}$ ergs.
The total nuclear energy output ($\sim4 \times 10^{54}$ ergs) producing the 
lobes in NGC 4472 is small compared to that seen in other galaxies,
particularly  those at cluster centers,
which possess nuclear energy outputs of up to $10^{61}$ ergs \citep[]{A2597}.
The expansion of the radio lobes in NGC 4472 evacuates the 
X-ray emitting gas within the lobes, but it does not compress the X-ray lobes 
enough to shock the surrounding gas or to produce bright rims like those
observed in galaxies with larger nuclear energy outputs \citep[]{A2597}.
From models by \citet[]{bubblemod}, we find a lobe rise velocity of 
$\sim$320 km s$^{-1}$.  
This rise velocity is approximately the sound speed 
in the gas, supporting the interpretation of the rise of the lobe as subsonic.

Larger scale structures in the halo gas are a product of the
interaction of that gas with the Virgo ICM.  While the radially
averaged emission fits a $\beta$ model surface brightness profile with
$\beta$=0.440$\pm$0.001 and core radius a=0.280$\pm$0.001 kpc, there
are significant deviations from radial symmetry.  The temperature
increases monotonically from 0.8 keV to 1.1 keV radially outward to 10
kpc from the nucleus, then rises to 1.4 keV along the NE edge of the
galaxy 10-12 kpc from the nucleus.  At this NE edge, the surface
brightness drops by more than an order of magnitude over a scale of
50\arcsec~(3.9 kpc).  The NE edge is the result of compression as NGC
4472 falls inward towards M87 through the Virgo ICM.

A tail-like structure in the X-ray emitting gas extends $\sim$8$'$ (36
kpc in the plane of the sky) to the SW of the galaxy core in the Chandra observation.
We find a temperature of 1.1 keV for the tail gas and a temperature of 1.4
keV south of the tail.  Surface brightness increases by a factor of 3
on the tail compared to emission from the galaxy halo at the same
radius.  The southern edge of the tail is very sharp, with surface
brightness dropping by a factor of 3 over an angular length of
$<$20\arcsec~(155 pc.)  The lower temperature SW tail was drawn out
from the bright central region of the halo by ram pressure.  Its sharp
edge is a line of sight effect, since much of the tail lies radially to
the observer.  We find a total tail length of 96 kpc and a total tail
mass of 1.05$\times$10$^7$ M$_{\odot}$ for the tail observed by
Chandra. From the ROSAT PSPC observation, we can trace the tail to
$\sim$11$'$, corresponding to a length on the sky of 133 kpc.

We also confirm emission from the nucleus (first observed by \citet[]{4472_nuc2})
and identify two luminous sources within $10''$ of  the nucleus as extended.
The two small extended sources 
sources have luminosities of $\sim7 \times $10$^{38}$ erg s$^{-1}$ 
and have no radio or optical
counterparts.  Modeling the two small extended sources as hot X-ray emitting
gas clouds, we find that they would be severely overpressurized relative to 
the ambient medium.  Thus, it is unlikely that these sources
are X-ray emitting gas clouds; instead, they are likely collections of X-ray 
point sources.

\acknowledgements

We would like to thank S. Borgani for insightful 
conversations.  We would also like to acknowledge 
J. Vrtilek and A. Vikhlinin for useful suggestions regarding data reduction.
The authors would also like to thank the anonymous referee for useful 
comments.
X-ray observations were obtained from the Chandra and HEASARC
archives, while radio observations were obtained from FIRST and HST
WFPC images from the STScI archive.  This work was supported by the
Smithsonian Institution, the Chandra Science Center and NASA contracts
NAS8-03060 and NAS8-01130.

\clearpage

\clearpage

\begin{thebibliography}{}
\bibitem[Becker et al.(2003)]{FIRST} Becker, R.H., Helfand, D.J., White, R.L., Gregg, M.D., \& Laurent-Muehleisen, S.A.  2003, http://sundog.stsci.edu
\bibitem[Birzan et al.(2004)]{syscav} Birzan, L., Rafferty, D.A., McNamara, B.R., Wise, M.W., \& Nulsen, P.E.J.  2004, \apj, accepted
\bibitem[Blakeslee et al.(2001)]{velocities} Blakeslee, J.P., Lucey,
J.R., Barris, B.J., Hudson, M.J., \& Tonry, J.L. 2001, \mnras, 327, 1004
\bibitem[Blanton et al.(2001)]{A2052} Blanton, E.L., Sarazin, C.L.,
McNamara, B.R., \& Wise, M. 2001, \apj, 558, L15
\bibitem[B\"ohringer et al.(1993)]{N1275a} B\"ohringer, H., Voges, W.,
Fabian, A.C., Edge, A.C., \& Neumann, D.M. 1993, \mnras, 264, L25
\bibitem[Borgani (private communication)]{grav_model} Borgani,
S. private communication.
\bibitem[Boroson $\&$ Thompson(1991)]{4472nuc4} Boroson, \& Thompson,  .  1991, \aj, 101, 111
\bibitem[Caon et al.(2000)]{star_orbit} Caon, N., Macchetto, D., \&
Pastoriza, M. 2000, \apjs, 127, 39
\bibitem[Churazov et al.(2000)]{N1275} Churazov, E., Forman, F.,
Jones, C., \& B\"ohringer, H. 2000, \aap, 356, 788
\bibitem[Dickey $\&$ Lockman(1990)]{nH} Dickey, J.M., \& Lockman,
F.J. 1990, \araa, 28, 215
\bibitem[Dosaj et al.(2002)]{N1404} Dosaj, A., Jones, C., Forman, W.R., Markevitch, M.L., \& Vikhlinin, A.A. 2002, \baas, 200, 4317
\bibitem[Elmegreen et al.(2000)]{M86_dust} Elmegreen, D.M., Elmegreen,
B.G., Chromey, F.R., \& Fine, M.S. 2000, \apj, 120, 733
\bibitem[Ensslin $\&$ Heinz(2002)]{bubblemod} Ensslin, T.A., \& Heinz,
S. 2002, \aap, 384, L27
\bibitem[Fabian et al.(2003)]{N1275b} Fabian et al.
2003, MNRAS, 344, L43
\bibitem[Finoguenov $\&$ Jones(2000)]{asca} Finoguenov, A., \& Jones,
C. 2000, \apj, 539, 603
\bibitem[Finoguenov $\&$ Jones(2001)]{M84} Finoguenov, A., \& Jones,
C. 2001, \apj, 547, L107
\bibitem[Forman, Jones, $\&$ Tucker(1985)]{einstein} Forman, W.,
Jones, C., \& Tucker, W. 1985, \apj, 293, 102
\bibitem[Forman et al.(2001)]{M86} Forman, W., Markevitch, M., 
Jones, C., Vikhlinin, A., \& Churazov, E. 2001,  
http://xxx.lanl.gov/ps/astro-ph/0110087
\bibitem[Forman et al.(1993)]{cor2} Forman, W., Jones, C., David, L.,
Franx, M., Makishima, K., \& Ohashi, T. 1993, \apj, 418, L55
\bibitem[Heinz et al.(2002)]{A4059} Heinz, S., Choi, Y.Y., Reynolds, C.S.,
\& Begelman, M.C. 2002, \apj, 569, L79 
\bibitem[Heinz et al.(2003)]{cfsim} Heinz et al. 2003.
\bibitem[Irwin $\&$ Sarazin(1996)]{ROSAT} Irwin, J.A., \& Sarazin,
C.L. 1996, \apj, 471, 683 
\bibitem[Jones et al.(2001)]{N4636} Jones, C., Forman, W., Vikhlinin,
A., Markevitch, M., David, L., Warmflash, A., Murray, S., \& Nulsen,
P.E.J. 2001, \apj, 567, L115
\bibitem[Kraft et al.(2003a)]{N507} Kraft, R.P., Laslo, N., Forman, W.R., Murray, S.S., Jones, C., Markevitch, M., Vikhlinin, A., \& Churazov, E.  2003, HEAD, 35, 1315
\bibitem[Kraft et al.(2003b)]{N7619} Kraft et al.  2003, in preparation.
\bibitem[Loewenstein et al.(2001)]{4472_nuc} Loewenstein, M.,
Mushotzky, R.F., Angelini, L., Arnaud, K., \& Quataert, E. 2001, \apj,
555, L21
\bibitem[Maccarone et al.(2003)]{4472_gc} Maccarone, T.J., Kundu, A., \& Zepf, S.E.  2003, \apj, 586, 814
\bibitem[Machecek et al.(2004)]{N1404b} Machecek et al. 2004, in preparation.
\bibitem[Markevitch(2000)]{cleaning} M. Markevitch. 2000,\\ http://asc.harvard.edu/cal/Links/Acis/acis/Cal\_prods/bkgrnd/current/index.html
\bibitem[Markevitch et al.(2000)]{A2142} Markevitch, M., Ponman, T. J., Nulsen, P. E. J., Bautz, M. W., Burke, D. J., David, L. P., Davis, D., Donnelly, R. H., Forman, W. R., Jones, C., Kaastra, J., Kellogg, E., Kim, D.-W., Kolodziejczak, J., Mazzotta, P., Pagliaro, A., Patel, S., Van Speybroeck, L., Vikhlinin, A., Vrtilek, J., Wise, M., \& Zhao, P.  2000, \apj, 541, 542
\bibitem[Markevitch et al.(2002)]{cf2} Markevitch, M., Gonzalez, A.H., David, L., Vikhlinin, A., Murray, S., Forman, W., Jones, C., \& Tucker, W.  2002, \apj, 567, 27
\bibitem[Matsumoto et al.(1997)]{cor4} Matsumoto, H., Koyama, K.,
Awaki, H., Tsuru, T., Loewenstein, M., \& Matsushita, K. 1997, \apj,
482, 133
\bibitem[McNamara et al.(1994)]{UGC76362} McNamara, B.R., Sancisi, R.,
Henning, P.A., \& Junor, W.  1994, \aj, 108, 844
\bibitem[McNamara et al.(2000)]{Hydra} McNamara, B.R., Wise, M.,
Nulsen, P.E.J., David, L.P., Sarazin, C.L., Bautz, M., Markevitch, M.,
Vikhlinin, A., Forman, W.R., Jones, C., \& Harris, D.E. 2000, \apj,
534, L135
\bibitem[McNamara et al.(2001)]{A2597} McNamara, B.R., Wise, M.,
Nulsen, P.E.J., David, L.P., Carilli, C.L., Sarazin, C.L., O'Dea,
C.P., Houck, J., Donahue, M., Baum, S., Voit, M., O'Connell, R.W.,
$\&$ Koekemoer, A. 2001, \apj, 562, L149
\bibitem[Nagar et al.(2002)]{4472_nuc3} Nagar, N.M., Falcke, H., Wilson, A.S., \& Ulvestad, J.S.  2002, \aap, 392, 53
\bibitem[Patterson $\&$ Thuan(1992)]{UGC76361} Patterson, R.J. \& Thuan, T.X.
1992, \apj, 400, L55
\bibitem[Serlemitsos et al.(1993)]{cor1} Serlemitsos, P.J.,
Loewenstein, M., Mushotzky, R.F., Marshall, F.E., \& Petre, R. 1993,
\apj, 413, 518
\bibitem[Statler \& McNamara (2002)]{NGC1700} Statler, T. \& McNamara, B., 2002, \apj, 581, 1032
\bibitem[Soldatenkov, Vikhlinin, \& Pavlinsky(2003)]{4472_nuc2} Soldatenkov, D.A., Vikhlinin, A.A., \& Pavlinsky, M.N.  2003, Astronomy Letters, 29, 298
Vikhlinin, and Pavlinsky (2003)
\bibitem[Sun et al.(2003)]{A478} Sun, M., Jones, C., Murray, S.S., Allen, S.W., Fabian, A.C., \& Edge, A.C. 2003, \apj, 587, 619
\bibitem[Thomas(1986)]{einstein2} Thomas, P.A. 1986, \mnras, 220, 949
\bibitem[Tonry et al.(2001)]{distance} Tonry, J.L., Dressler, A.,
Blakeslee, J.P., Ajhar, E.A., Fletcher, A.B., Luppino, G.A., Metzger,
M.R., \& Moore, C.B. 2001, \apj, 546, 681
\bibitem[Vikhlinin et al.(2001)]{cf1} Vikhlinin, A., Markevitch, M., \& Murray, S.S. 2001, \apj, 551, 160
\bibitem[Volmer et al.(2001)]{ram_model}Volmer, B., Cayatte, V.,
Balkowski, C., \& Duschl, W.J. 2001, \apj, 561, 708
\end{thebibliography}
\end{document}